\documentclass[useAMS,usenatbib]{mn2e}

\usepackage{graphicx}                   
\input{epsf.sty}                        
\input{psfig.sty}
\begin{document}

\title[Superflares: giant quakes in quark stars?]
{The superflares of soft $\gamma$-ray repeatres: giant quakes in
solid quark stars?}

\author[Xu, Tao, Yang]{R. X. Xu$^{1}$, D. J. Tao$^{2}$, and Y. Yang$^{1}$\\
$^{1}$School of Physics, Peking University, Beijing 100871,
China\\
$^{2}$School of Mathematical science, Peking University, Beijing
100871, China}

\maketitle

\begin{abstract}
Three times of supergiant flares from soft $\gamma$-ray repeatres
are observed, with typical released energy of $\sim 10^{44-47}$
erg. A conventional model (i.e., the magnetar model) for such
events is catastrophic magnetism-powered instability through
magnetohydrodynamic process, in which a significant part of
short-hard $\gamma$-ray bursts could also be the results of
magnetars.
Based on various observational features (e.g., precession, glitch,
thermal photon emission) and the underlying theory of strong
interaction (quantum chromodynamics, QCD), it could not be ruled
out yet that pulsar-like stars might be actually solid quark
stars.
Strain energy develops during a solid star's life, and starquakes
could occur when stellar stresses reach a critical value, with
huge energy released.
An alternative model for supergiant flares of soft $\gamma$-ray
repeatres is presented, in which energy release during a star
quake of solid quark stars is calculated.
Numerical results for spherically asymmetric solid stars show that
the released gravitational energy during a giant quake could be as
high as $10^{48}$ erg if the tangential pressure is slightly
higher than the radial one.
Difficulties in magnetar models may be overcome if AXPs/SGRs are
accreting solid quark stars with mass $\sim (1-2)M_\odot$.
\end{abstract}


\section{Introduction}

Pulsar-like stars keep to manifest surprising observational
features since their first discovery in 1967.
One of their extraordinary behaviors, whose origin and
astrophysical implications are hotly debated about, is the
superflare from soft $\gamma$-ray repeaters (SGRs).
Only three such events have been detected in three of the four
SGRs: 1979/03/05 of SGR 0525-66 (spin period $P=8.1$ s),
1998/08/27 of SRG 1900+14 ($P=5.16$ s), 2004/12/27 of SGR 1806-20
($P=7.45$ s). The released energy in the former two flares could
be $\sim 10^{44\sim 45}$ erg, while $\sim 10^{47}$ erg in the
third.
A peculiar nature of superflare is the initial brief ($\sim 0.2$
s) spikes of $\gamma-$rays with energies up to several MeV, which
contain most of the flare energy and are followed by tails lasting
minutes \cite[e.g.,][]{Hurly05}.
Additionally, quasi-periodic oscillations (QPOs) during
superflares were found soon after the onsets of the superflares in
SGR 1806-20 \cite[at frequency $f\sim 93$ Hz, ][]{Israel05}, in
SGR 1900+14 \cite[$f\sim 84$ Hz, ][]{sw05}, and in SGR 0526-66
\cite[$f\sim 43$ Hz, ][]{Barat83}.
Higher frequency oscillations at about 150, 625, and 1,840 Hz are
also detected from the superflare of SGR 1806-20 \citep{sw06}.

Current models for the superflares are in the scenario of
magnetars, a kind of neutron stars with polar magnetic fields in
the range $10^{14}\sim 10^{15}$ G, including the SGRs and the
anomalous X-ray pulsars (AXPs). The quiescent X-ray emission with
luminosity $\sim 10^{34\sim 36}$ erg/s as well as the superflares
of SGRs are supposed to be powered by magnetic field decay
\cite[e.g.,][]{wt05}.
However, there are still some debates on magentars (see \S3 for
more discussion on their existence), although they are really
popular in the astrophysical society.
Alternatively, we propose here that a giant quake in a solid quark
star may result in a superflares and reproduce the general
observational features.

A quark star is composed dominantly  by quark matter which is a
direct consequence of the asymptotic freedom nature in quantum
chromodynamics (QCD), the underlying theory believed for the
elementary strong interaction.
It is worth noting that this stellar quark matter at low
temperature should be very different from the hot quark matter to
be searched in relativistic heavy ion colliders.
A degenerate Fermi gas of quarks is expected at extremely high
density and temperature, while Cooper pairing of quarks near the
Fermi surface occurs in cool and dense (but not asymptotically)
quark matter because of the strong and attractive QCD quark-quark
interaction. A condensate of the pairs may then result in a color
superconductivity (CSC) phase in this case \cite[ and references
therein]{ar06}.
However, a solid state with quark-clusters in periodic lattices
was also conjectured in a parametric region where the density
could be lower than that of CSC state \citep{xu03,xu05}. This
hypothetic state could still not be ruled out by simple QCD
principles as well as astrophysical observations \cite[ and
references therein]{xu06a}.
Unfortunately, due to the non-linear nature of QCD, one can {\em
not} now obtain with certainty the critical parameters (e.g.,
baryon density and temperature) for the asymptotically free quark
phase, the CSC phase, and the (solid) quark-cluster phase.

Pulsar-like stars, including SGRs and AXPs, are compact remnants
of evolved stars, the nature of which is still a matter of
controversy. Though these stars are popularly thought to be normal
neutron stars, no convincing work, either theoretical from first
principles or observational, has excluded the possibility that
they are actually quark stars.
Besides the rotational, thermal, magnetic, and
accretion-gravitational ones, the remaining free energy for a
solid quark star includes additional elastic and
quake-gravitational energies \citep{Horvath05,xu06}.
In this paper, we will show that these additional free energies of
solid quark stars could be high enough to power the superflares of
SGRs.
Though the energy budget quest of the superflares is focused on,
the radiative mechanisms as well as astrophysical implications of
such events are discussed too.

\section{The model}

It is very fundamental to study static and spherically symmetric
gravitational sources in general relativity, especially for the
interior solutions.
The Tolman-Oppenheimer-Volkoff (TOV) solution is only for perfect
fluid. However, for solid quark stars, since the local press could
be {\em  anisotropic} in elastic matter, the radial pressure
gradient could be partially balanced by the tangential shear force
although a general understanding of relativistic, elastic bodies
has unfortunately not been achieved \citep[e.g.,][]{ks04}.
The origin of this local anisotropic force in solid quark stars
could be from the development of elastic energy as a star (i)
spins down (its ellipticity decreases) and (ii) cools (it may
shrink).
Release of the elastic as well as the gravitational energies would
be not negligible, and may have significant astrophysical
implications.

Let's numerically calculate the structure of solid quark stars as
following. For the sack of simplicity, we only deal with
spherically symmetric sources in order to make sense of possible
astrophysical consequence of solid quark stars.
By introducing respectively radial and tangential pressures, $P$
and $P_\bot$, the stellar equilibrium equation of static
anisotropic matter in Newtonian gravity is \cite[ Eq. (2.4)
there]{hs97}: ${\rm d}P/{\rm d}r=-Gm(r)\rho/r^2+2(P_\bot-P)/r$,
where $\rho$, $m(r)$ and $G$ denote, respectively, mass density,
mass interior the radius $r$ and the gravitational constant.
However, in Einstein's gravity, this equilibrium equation is
modified \citep[e.g.,][]{liu99},
\begin{eqnarray}
\frac{{\rm d}P}{{\rm d}r} = -\frac{Gm(r)\rho}{r^2}\frac{(1 +
\frac{P}{\rho c^2})(1 + \frac{4\pi r^3P}{m(r)c^2})}{1 -
\frac{2Gm(r)}{rc^2}} +
\frac{2\epsilon}{r}P,%
\label{P'}
\end{eqnarray}
where $P_\bot=(1+\epsilon)P$ is introduced. In case of isotropic
pressure, $\epsilon=0$, Eq.(\ref{P'}) turns out to be the TOV
equation.
It is evident from Eq.(\ref{P'}) that the radial pressure
gradient, $|{\rm d}P/{\rm d}r|$, decreases if $P_\bot>P$, which
may result in a higher maximum mass of compact stars. Whether the
$2.1M_\odot$-millisecond pulsar \citep{nice05} in a binary system
with a helium white dwarf secondary is relevant to this nature of
solid stars could be an interesting topic.
One can also see that a sudden decrease of $P_\bot$ in a star may
cause substantial energy release, since the star's radius
decreases and the absolute gravitational energy increases.

A quark star would initially be in a fluid state well approximated
by a perfect fluid. In a simplified version of the bag model for
quark matter, the equation of state for fluid quark matter could
be
\begin{eqnarray}
P = {1\over 3}(\rho - 4B) c^2,%
\label{eos}
\end{eqnarray}
where the bag constant $B\cdot c^2$ could be between 60 MeV/fm$^3$
and 110 MeV/fm$^3$. Since no equation of sate for solid quark
matter is available, we may just approximate the radial pressure
by Eq.(\ref{eos}) in the following calculations.
Also the equation below holds,
\begin{equation}
\frac{{\rm d}m(r)}{dr} = 4\pi r^2 \rho.%
\label{mr}
\end{equation}
We have to integrate numerically these differential equations of
Eq.(\ref{P'}-\ref{mr}), which are complete for $P$, $\rho$ and
$m(r)$ if $B$ and $\epsilon$ are certain, in order to know the
$\epsilon$-dependent global structure (e.g., radius and mass) of a
solid quark star.

Substituting Eq. (\ref{eos}) into Eq.(\ref{mr}) and Eq.
(\ref{P'}), we have
\begin{equation}
\left\{
\begin{array}{l}
\frac{dm(r)}{dr} = 4\pi r^2 (4B + \frac{3P}{c^2}),\\
\frac{dP}{dr} = -\frac{Gm(r)(4B + \frac{3P}{c^2})}{r^2}\frac{(1 +
\frac{P}{(4B + \frac{3P}{c^2}) c^2})(1 + \frac{4\pi
r^3P}{m(r)c^2})}{1 - \frac{2Gm(r)}{rc^2}} + \frac{2\epsilon}{r}P.%
\end{array}\right.%
\label{num-eq}
\end{equation}
The precision of numerical solution should be very {\em high} in
order to obtain the small-$\epsilon$-dependent features from
Eq.(\ref{num-eq}).
Initially, the core of a star is supposed to be homogeneous, with
the density $\rho=\rho_0$, the radius $r=0.1$ cm, $m(0.1) =
4\rho_0\pi r^3 / 3$, and $P(0.1) = (\rho_0 - 4B)c^2/3$ in the
computation.
Numerical method can then be used to integrate Eq.(\ref{num-eq})
from $r=0.1$ cm to the boundary, $r=R$, of the star, step by step
(with index $n$).
At the star's boundary, $P(R) = 0$ and $M=m(R)$, with $R=r_n$ if
and only if $P(R_n) \geq 0 , P(R_{n+1}) < 0$.
Approximated in a flat spacetime, the total gravitational energy
$E$ and the stellar moment of inertia $I$ could also been obtained
in the numerical process, by integrating the following equations,
\begin{eqnarray}
{\rm d}E &=& - \frac{Gm(r)}{r}{\rm d}m(r) = -\frac{Gm(r)}{r}(4\pi
r^2 \rho
{\rm d}r)\nonumber\\
&=& -4\pi (\frac{3P}{c^2} + 4B)Gm(r)r{\rm d}r,\\
{\rm d}I &=& \frac{2}{3}r^2{\rm d}m(r) = \frac{2}{3}r^2(4\pi r ^ 2
\rho {\rm d}r)\nonumber\\
&=& \frac{8\pi}{3}(\frac{3P}{c^2} + 4B)r^4{\rm d}r.
\end{eqnarray}
The Runge-Kutta method of order-4 is used in the code.

Some issues are worth noting during the numerical process.
It is very necessary to improve the precision in order to obtain
the differences, which we are interested, of $E(\epsilon)$ and
$I(\epsilon)$ by numerical integration, since both the related
numbers and the domains in this problem vary in 7 orders of
magnitudes.
Computing error comes mainly from the decision of a star's
boundary. To improve the precision of integration, we have to
divide the domain to be smaller and smaller near the boundary,
until the quantities calculated are credible.

\begin{figure}
\includegraphics[width=3in]{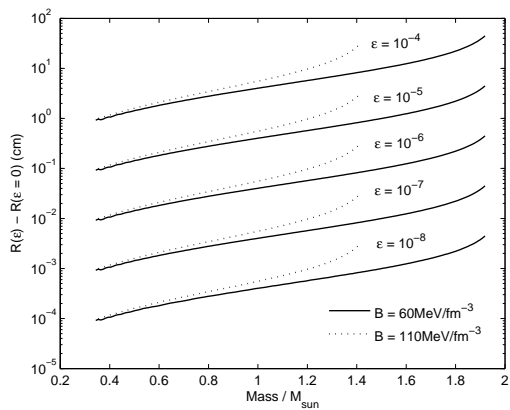}%
\caption{%
The difference of stellar radius as a function of stellar mass.
Solid lines are for bag constant $B=60$ MeV/fm$^3$, and dashed
lines for $B=110$ MeV/fm$^3$. The quantity of $\epsilon$ is
defined by introducing $P_\bot=(1+\epsilon)P$, where $P_\bot$, $P$
are the radial and tangential pressures, respectively.
\label{R}}
\end{figure}

\begin{figure}
\includegraphics[width=3in]{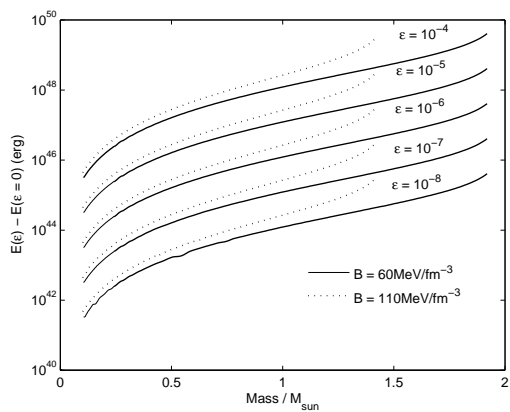}%
\caption{%
Same as in Fig.1, but for the difference of gravitational energy.
\label{E}}
\end{figure}

\begin{figure}
\includegraphics[width=3in]{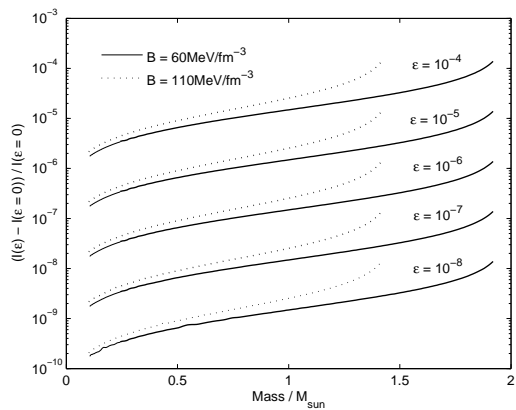}%
\caption{%
Same as in Fig.1, but for the ratio difference of moment of
inertia.
\label{I}}
\end{figure}

Our calculation results are shown in Fig.\ref{R} to Fig.\ref{I}.
Starquakes may result in a sudden change of $\epsilon$, with an
energy release of the gravitational energy as well as the
tangential strain energy.
Generally, it is evident that the differences of radius,
gravitational energy, and moment of inertia increase
proportionally to stellar mass and the parameter $\epsilon$. This
means that an event should be more important for a bigger change
of $\epsilon$ in a quark star with higher mass.
It is shown in Fig.\ref{R} that the stellar radius varies
insignificantly when small $\epsilon$ is considered
($R(\epsilon)-R(0)$ is only from $10^{-4}$ cm $\sim 10$ cm for
$\epsilon=10^{-8\sim -4}$ if stellar mass is $\sim M_\odot$).
Only gravitational energy release is considered in Fig.\ref{E},
which could be as high as $10^{48}$ erg if $\epsilon\sim 10^{-4}$
and $M\sim M_\odot$. Typical energy of $10^{44\sim 47}$ erg is
released during superflares of SGRs, and we may then propose a
giant starquake with $\epsilon\la 10^{-4}$ could produce such a
flare.
A sudden change of $\epsilon$ can also result in a jump of spin
frequency, $\Delta \Omega/\Omega=-\Delta I/I$. We may expect from
Fig.\ref{I} that glitches with $\Delta \Omega/\Omega\sim
10^{-10\sim -4}$ could occur for parameters of $M=(0.1\sim
1.4)M_\odot$ and $\epsilon=10^{-9\sim -4}$.
It is suggestive that a giantflare may accompany a high-amplitude
glitch.

\section{Conclusions and Discussions}

We suggest alternatively that the superflares of soft $\gamma$-ray
repeaters could be the results of giant quakes of solid quark
stars. Numerical calculations for spherically asymmetric solid
stars show that the released gravitational energy during a giant
quake could be as high as $10^{48}$ erg if the tangential pressure
is slightly higher (only $\sim (1+10^{-5})$ times) than the radial
one in a star with mass $\sim M_\odot$.
However, a detail process by which the released energy is
transformed into radiation is still not clear during a quark.
The quake-calculation presented in this paper is different from
that of \cite{z04}.

A direct consequence in the model here is that, besides giant
quakes, much small quakes especially of aged quark stars (both
their magnetospheric and thermal radiations would be low enough
not to be detected regularly by recent facilities) with low masses
are expected to occur as hard X-ray transients.
There is a rough tendency that both the jump amplitudes and the
frequency of the glitches are apparently decreased with the pulse
period when the pulsars become old \citep{l00}. No glitch pulsar
with the period longer than 0.7 second has been detected.
According to the law of seismology, no big quake occurs if many
small ones happen frequently, but a giant quake may take place
after a long time of silence (i.e., no quake period).
If this law applies, an old solid quark star with long spin period
($\gg 1$ s), which may have enough time to leave its host galaxy,
could have a big quake (crash) after a long time of silence.
Part of short $\gamma$-ray burst could probably rare flares of
middle or low amplitudes of starquakes in the galactic halo.
More such events could be recorded by the Swift or the future HXMT
(hard X-ray modulation telescope).

Anomalous X-ray pulsars/Soft $\gamma$-ray repeaters (AXPs/SGRs)
are supposed to be magnetars. But an alternative suggestion is
that they are normal-field pulsar-like stars which are in an
accretion propeller phase \citep{Alpar01,chn00}. The very
difficulty in the later view point is to reproduce the irregular
bursts, even with peak luminosity $\sim 10^7L_{\rm Edd}$ (SGR
0526-66; $L_{\rm Edd}$ the Eddington luminosity).
Both the possibilities of the bombardments of comet-like objects
\citep[e.g., strange planets,][]{xu05} to bare strange stars and
of giant quakes in solid quark stars could remove the difficulty
during super-bursts since the interaction in quark matter should
be very strong.
Based on the calculation in Fig.\ref{E}, it is conjectured that
SGRs/AXPs could be quark stars with masses in order of $M_\odot$,
since more energy would be released during their quakes as bursts.
Other pulsar-like stars (compact center objects and dim thermal
``neutron stars''), which have small bolometric radii and low
surface temperatures, could be quark stars with lower masses.
Actually, there could be observations as well as theoretical
arguments which do not favor the magnetar idea.
(i). The superstrong field of magnetars are supposed to be created
by MHD-dynamo action of rapid rotating protoneutron stars with
spin period $<3$ ms. The Poynting flux and the relativistic
particle ejection of such a star should power effectively the
supernova remnants. Such energetic remnants are expected in
magnetar models, but had not been detected \citep{vk06}.
(ii). Dust emission around pulsar-like stars (e.g., AXPs) was
proposed to test observationally the propeller scenarios of quark
stars with Spitzer or SCUBA \citep{xu05,xu06}. Actually a recent
discovery of mid-infrared emission from a cool disk around an
isolated young X-ray pulsar, 4U 0142+61, is reported \citep{wck06}
although it is still a matter of debate whether significant
propeller torque of fallback matter acts on the center star.
(iii) The pressure should be very anisotropic in a relativistic
degenerate neutron gas in equilibrium with a background of
electrons and protons when the magnetic field is stronger than the
critical field. The vanishing of the equatorial pressure of the
gas would result in a transverse collapse, and a stable magnetar
could then be unlikely \citep{mrs03}.
(iv). It is still a matter of debate whether the absorption
features in SGR 1806-20 can be interpreted by proton or electron
cyclotron resonance \citep{xwq03}. The field is only $\sim 5\times
10^{11}$ G in the context of electron cyclotron lines.
(v). \cite{mm04} suggested that the persistent X-ray emission of
AXPs/SGRs could be originated from the cyclotron mechanism acting
near the surface of a star with normal field $\sim 10^{12}$ G,
while short-time-scale cataclysmic events on the neutron star
could lead to the bursts.
Considering these criticisms, we think that alternative ideas for
understanding AXPs/SGRs phenomena are very necessary.

How to test this starquake model and the magnetar model for
AXPs/SGRs?
Because of the change of mass momentum, gravitational wave
radiation may accompany a giant starquake of a solid quark star,
but such radiation may not be significant in the magnetar model,
at least the timescale and other feature of the waves should be
very different.
The observed QPOs could also hint the nature of superflares of
SGRs. A detail investigation on stellar torsional vibration and a
comparison between the QPO-characters in starquake and magnetar
models \citep[e.g., ][]{gsa06,bt06} are very necessary.
In our model for SGRs, glitches may occur during gravity-induced
superflares. This could be applied to distinguish the models.
Interestingly, an glitch of $\Delta \nu/\nu=4.2\times 10^{-6}$ was
discovered in 1E 2259+586, which preceded the the 2002 out-burst
activity \citep{woods04}. Actually, occasional 4 glitches had been
detected in 3 AXPs; sometimes they were associated with radiative
events \citep{kaspi06}.
However, only gravitational energy change is calculated in this
paper. It is worth noting that stress energy developed in solid
quark stars should be another kind of free energy. No significant
frequency glitch occurs if a superflare originates from the
release of stress energy.
How much is the stress energy during the evolution of a solid
quark star? Is it comparable to the gravitational energy? These
are questions to be answered in the future.

A pulsar-like star could be monopole-charged electrically, due to
either the global structure of current flows in pulsar
magnetospheres \citep{xcq06} or maybe other reasons.
\cite{weber06} showed that, depending of the amount of electric
charge, the star's structure and specifically the mass-radius
relationship might be drastically modified.
A loss of the electricity could also increase the parameter
$\epsilon$ (i.e., the tangential pressure becomes higher and
higher than the radial one as the electricity decreases). A
starquake would occur as the star discharges, which we have not
included in calculations.

{\em Astrophysical links between AXPs and SGRs?}
The persistent X-ray emission from AXPs/SGRs are alternatively
suggested to be accretion-powered \cite[i.e., accretor with
conventional magnetic fields, ][]{Alpar01}. But how to reproduce
naturally SGR-like bursts in this scenario?
Quark stars with low masses ($\ll M_\odot$) are self-confined by
the strong interactions between quarks, whereas
gravitation-binding can not be negligible for quark stars with
much higher masses ($\sim M_\odot$). This results in an
approximate relation of $M\propto R^3$ for low masses but
violation for higher masses in the mass($M$)-radius($R$) diagram.
It is well known that the stellar radius decreases as the mass
increases for pure gravitation-confined Fermion stars (e.g., white
dwarfs with state of perfect electron gases).
An accreting solid quark star may undergo blazing quakes if it has
a mass of $\ga M_\odot$ when gravitation-binding dominates, since,
in this case, the gravitation-induced shear force becomes stronger
and stronger as the star accretes.
Therefore, it is proposed \citep{xu06} that SGRs/AXPs might be
solid quark stars with $\sim (1-2)M_\odot$, while other
pulsar-like stars could be of low-mass.
In order to make sense about the values of mass and radius of
gravitation-dominated quark stars, we calculate these two
parameters for fluid quark stars, using simply an equation of
state of Eq.(\ref{E}). It is shown in Fig. \ref{mr} that the
difference between the mass of quark stars with maximum mass
(${\rm d}M/{\rm d}R=0$) and that with maximum radius (${\rm
d}R/{\rm d}M=0$) could be $\Delta M\sim 0.1M_\odot$ for possible
bag constant ($60-110$) MeV$\cdot$fm$^3$. The corresponding radius
difference is about $\Delta R \sim 0.4$ km.
The maximum gravitational-energy release during successive quakes
of an accreting solid quark star from the point of ``${\rm
d}R/{\rm d}M=0$'' to the point of ``${\rm d}M/{\rm d}R=0$'' could
typically be $E_{\rm quake}\sim G({M(M+\Delta M)\over R-\Delta
R}-{M^2\over R})\sim 10^{52}$ erg.
But the total energy release for the persistent X-ray emission is
$E_{\rm persistent}\sim GM\Delta M/R\sim 10^{52}$ erg.
For AXPs/SGRs with persistent X-ray luminosity $L_{\rm x}\sim
10^{35}$ erg/s, the lifetime of such sources could be $E_{\rm
persistent}/L_{\rm x}\sim 10^{10}$ yrs. This means that it would
cost nearly the Hubble time to increase the stellar mass to the
maximum value (so that the star may collapse to a black hole) {\em
if} the accretion rate is a constant.
\begin{figure}
\includegraphics[width=3in]{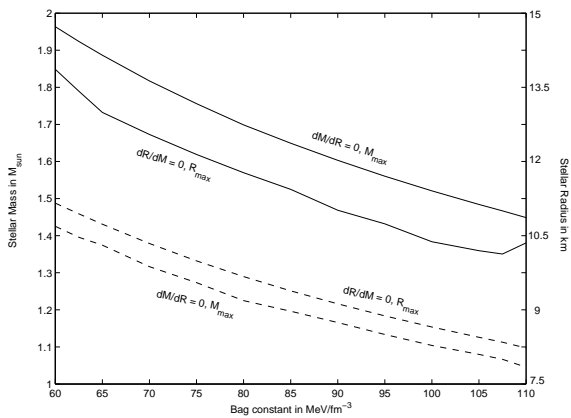}%
\caption{%
An illustration of stellar mass (solid lines, in $M_\odot$) and
radius (dashed lines, in km) of gravitation-dominated quark stars,
as a function of the bag constant. The masses as well as the radii
of quark stars with maximum radius (${\rm d}R/{\rm d}M=0$) or
maximum mass (${\rm d}M/{\rm d}R=0$) are shown. It is evident that
the mass and radius differences could be about $0.1M_\odot$ and
0.4 km, respectively.
\label{mr}}
\end{figure}

{\em Radiative mechanisms of star-quake-induced flares.}
Because of the starquakes of neutron stars, self-induction
electric field is created \citep{tlk02,bt06}. The strong electric
field could initiate avalanches of pair creation in the
magnetosphere and certainly accelerate particles, resulting in
high energy bursts observed.
Similar mechanism would also work when starquakes occur in solid
quark stars. The only difference should be that no ions can be
supplied from the surface of bare quark stars (i.e.,
lepton-dominated plasma forms above quark surfaces).
The energy of oscillations excited by starquakes could be
transported to the plasma corona by, e.g., Alfv\'en waves, being
similar to the case of heating the solar wind \citep{t88}.
It is interesting to know if such Alfvenic fluctuations,
originated from stellar torsional oscillations, could result in
the observed QPOs.
In the conventional magnetar model, the field should be at least
$B \simeq 8\times 10^{16}$ G in order to reproduce the superflare
of SRG 1806-20, if the efficiency of transforming magnetic energy
to the burst emission during the flare is $\sim 10^{-2}$. This low
limit means that magnetic field dominates (i.e., to control the
motion by magnetic field rather by crust) in the part of crust
with density $\rho < \rho_{\rm max}\sim 2.7\times 10^{11}$
g/cm$^3$. Note that $\rho_{\rm max}$ is in a same order of the
density of neutron drip $\rho_{\rm drip}=4.3\times 10^{11}$
g/cm$^3$.
It is possible (or not) that starquake could occur in the
out-crust (with density $\rho < \rho_{\rm drip}$) of normal
neutron stars, but the disadvantage is that the shear modulus
\citep{Fuchs26}, $\mu\propto Z^2$, in the deep crust might not be
high enough to crack, since the atomic charge $Z$ becomes smaller
and smaller in deeper crust due to neutronization.
However, for solid quark stars, starquakes could certainly exist
because of hight shear modulus as well as high density, even if
the stars have field as strong as $10^{16}$ G.

In conclusion, starquakes of solid quark stars are proposed for
the soft $\gamma$-ray superflares, with the persistent X-ray
emission to be fallback accretion-powered, for AXPs/SGRs.
Those problems in the magnetar models could be circumvented if
AXPs/SGRs are accreting solid quark stars with masses $\sim
(1-2)M_\odot$.
The rigidity of solid quark stars should be high enough for strong
quakes to occur, and the energy budget would not be a problem.


{\em Acknowledgments}:
RXX thanks Dr. Bing Zhang for a discussion of possible links
between AXPs and SGRs. This work is supported by NSFC (10573002)
and the Key Grant Project of Chinese Ministry of Education
(305001).


\end{document}